\documentclass[aps,prc,twocolumn,superscriptaddress,showpacs]{revtex4}
\usepackage[dvips]{graphicx}
\usepackage{amsmath,amssymb,times}
\usepackage{ulem}

\newcommand{\beq}{\begin{equation}}
\newcommand{\eeq}{\end{equation}}
\newcommand{\bea}{\begin{eqnarray}}
\newcommand{\eea}{\end{eqnarray}}

\DeclareSymbolFont{boldletters}{OML}{cmm} {b}{it}
\DeclareSymbolFontAlphabet{\mathbit}{boldletters}
\DeclareMathSymbol{\alpha}{\mathalpha}{letters}{"0B}
\DeclareMathSymbol{\beta}{\mathalpha}{letters}{"0C}
\DeclareMathSymbol{\gamma}{\mathalpha}{letters}{"0D}
\DeclareMathSymbol{\delta}{\mathalpha}{letters}{"0E}
\DeclareMathSymbol{\epsilon}{\mathalpha}{letters}{"0F}
\DeclareMathSymbol{\zeta}{\mathalpha}{letters}{"10}
\DeclareMathSymbol{\eta}{\mathalpha}{letters}{"11}
\DeclareMathSymbol{\theta}{\mathalpha}{letters}{"12}
\DeclareMathSymbol{\iota}{\mathalpha}{letters}{"13}
\DeclareMathSymbol{\kappa}{\mathalpha}{letters}{"14}
\DeclareMathSymbol{\lambda}{\mathalpha}{letters}{"15}
\DeclareMathSymbol{\mu}{\mathalpha}{letters}{"16}
\DeclareMathSymbol{\nu}{\mathalpha}{letters}{"17}
\DeclareMathSymbol{\xi}{\mathalpha}{letters}{"18}
\DeclareMathSymbol{\pi}{\mathalpha}{letters}{"19}
\DeclareMathSymbol{\rho}{\mathalpha}{letters}{"1A}
\DeclareMathSymbol{\sigma}{\mathalpha}{letters}{"1B}
\DeclareMathSymbol{\tau}{\mathalpha}{letters}{"1C}
\DeclareMathSymbol{\upsilon}{\mathalpha}{letters}{"1D}
\DeclareMathSymbol{\phi}{\mathalpha}{letters}{"1E}
\DeclareMathSymbol{\chi}{\mathalpha}{letters}{"1F}
\DeclareMathSymbol{\psi}{\mathalpha}{letters}{"20}
\DeclareMathSymbol{\omega}{\mathalpha}{letters}{"21}
\DeclareMathSymbol{\varepsilon}{\mathalpha}{letters}{"22}
\DeclareMathSymbol{\vartheta}{\mathalpha}{letters}{"23}
\DeclareMathSymbol{\varpi}{\mathalpha}{letters}{"24}
\DeclareMathSymbol{\varrho}{\mathalpha}{letters}{"25}
\DeclareMathSymbol{\varsigma}{\mathalpha}{letters}{"26}
\DeclareMathSymbol{\varphi}{\mathalpha}{letters}{"27}
\DeclareMathSymbol{\Gamma}{\mathalpha}{letters}{"00}
\DeclareMathSymbol{\Delta}{\mathalpha}{letters}{"01}
\DeclareMathSymbol{\Theta}{\mathalpha}{letters}{"02}
\DeclareMathSymbol{\Lambda}{\mathalpha}{letters}{"03}
\DeclareMathSymbol{\Xi}{\mathalpha}{letters}{"04}
\DeclareMathSymbol{\Pi}{\mathalpha}{letters}{"05}
\DeclareMathSymbol{\Sigma}{\mathalpha}{letters}{"06}
\DeclareMathSymbol{\Upsilon}{\mathalpha}{letters}{"07}
\DeclareMathSymbol{\Phi}{\mathalpha}{letters}{"08}
\DeclareMathSymbol{\Psi}{\mathalpha}{letters}{"09}
\DeclareMathSymbol{\Omega}{\mathalpha}{letters}{"0A}

\begin{document}

\title{Dual quark condensate in the Polyakov-loop extended Nambu--Jona-Lasinio model}

\author{Kouji Kashiwa}
\email[]{kashiwa@phys.kyushu-u.ac.jp}
\affiliation{Department of Physics, Graduate School of Sciences, Kyushu University,
             Fukuoka 812-8581, Japan}
             
\author{Hiroaki Kouno}
\email[]{kounoh@cc.saga-u.ac.jp}
\affiliation{Department of Physics, Saga University,
             Saga 840-8502, Japan}

\author{Masanobu Yahiro}
\email[]{yahiro@phys.kyushu-u.ac.jp}
\affiliation{Department of Physics, Graduate School of Sciences, Kyushu University,
             Fukuoka 812-8581, Japan}

\date{\today}

\begin{abstract}
The dual quark condensate $\Sigma^{(n)}$ proposed recently as a new order parameter of the spontaneous  breaking of the ${\mathbb Z}_3$ symmetry are evaluated by the Polyakov-loop extended Nambu--Jona-Lasinio model, where $n$ are winding numbers. 
The Polyakov-loop extended Nambu--Jona-Lasinio model well reproduces lattice QCD data on $\Sigma^{(1)}$ measured 
very lately. 
The dual quark condensate $\Sigma^{(n)}$ at higher temperatures is 
sensitive to the strength of the vector-type four-quark interaction 
in the Polyakov-loop extended Nambu--Jona-Lasinio model and hence a good quantity to determine the strength. 
\end{abstract}

\pacs{11.30.Rd, 12.40.-y}
\maketitle


Recently, a new order parameter of the $\mathbb{Z}_3$ center symmetry 
was proposed by using the chiral condensate $\sigma$ 
and evaluated by quenched \cite{Bilgici1,Bilgici2} and 
full lattice QCD~\cite{Bilgici3}; 
the new order parameter is called {\it the dual quark condensate}. 
This makes it possible 
to discuss the connection between the quark confinement 
and the chiral symmetry. 
The relation can be discussed 
also by the Polyakov-loop extended NJL (PNJL) model~\cite{Fukushima1} 
in which the Polyakov loop $\Phi$ is approximately 
treated as a classical variable.
Actually, extensive studies are made on the relation; 
for example, see Refs.~\cite{Ratti,Rossner,Costa,Fukushima1,
Kashiwa1,Abuki,Sakai,Sakai2,Hell,Kashiwa2, Kouno,Nam}. 
The dual quark condensate is calculated also in the color $SU(2)$ system 
by the Dyson-Schwinger equation~\cite{Fischer}, but this approach does not 
treat the confinement mechanism dynamically. 
Very recently, the functional renormalization-group method was also 
applied to evaluate the dual quark condensate in the color $SU(3)$ 
system~\cite{Braun}.

In this paper, we evaluate the dual quark condensate by using 
the PNJL model and show that the PNJL result can reproduce 
full lattice QCD (LQCD) data~\cite{Bilgici3} on the dual quark condensate.

We consider the quark field $q$ that obeys 
a twisted temporal boundary condition 
\begin{align}
q({\bf x}, \beta) = e^{-i \varphi} q({\bf x}, 0),
\label{eq:BC}
\end{align}
where $\varphi$ is a twisted angle. 
Now we define 
the $\varphi$-dependent chiral condensate~\cite{Bilgici1,Bilgici2,Bilgici3} by 
\bea
\sigma(\varphi)= - 
\frac{1}{V} \langle {\rm Tr}[(m+D_{\varphi})^{-1}] \rangle ,  
\label{sigma-phi}
\eea
where the twisted boundary condition \eqref{eq:BC} is imposed on 
the Dirac operator $D_{\varphi}$, 
but the bracket $\langle \dots \rangle$ keeps the anti-periodic boundary 
condition $\varphi=\pi$~\cite{Bilgici1,Bilgici2,Bilgici3}.
The dual quark condensate $\Sigma^{(n)}$ is defined by
\begin{align}
\Sigma^{(n)}
&= -\int_0^{2\pi} \frac{d\varphi}{2\pi} e^{-in\varphi} \sigma({\varphi}) 
\label{DPL}
\end{align}
with winding numbers $n$~\cite{Bilgici1,Bilgici2,Bilgici3}. 
In particular, $\Sigma^{(1)}$ is 
called {\it the dressed Polyakov loop}. The winding number is 1 in 
both $\Sigma^{(1)}$ and $\Phi$.
In full LQCD calculations of Ref~\cite{Bilgici3}, the dressed Polyakov-loop is evaluated with $m/T=0.032$.
We can expect that the two have similar $T$ dependence to each other. 
Thus, the dual quark condensate relates the chiral condensate to 
the quark confinement~\cite{Bilgici1,Bilgici2, Bilgici3, Fischer, Braun, Synatschke}.

We start with the two-flavor PNJL Lagrangian 
with the vector-type four-quark and the scalar-type eight-quark 
interactions; see Ref.~\cite{Kashiwa1} for the details.
The scalar-type eight-quark interaction makes 
the chiral phase transition stronger.
Hence, the PNJL result becomes consistent with LQCD data at finite $T$ 
\cite{Kashiwa1,Kashiwa3,Sakai2}. Furthermore, both 
the scalar-type eight-quark and the vector-type four-quark interaction 
are necessary for the PNJL model 
to reproduce LQCD data~\cite{FP,Elia,Chen} at imaginary chemical potential~\cite{Sakai2}.

 
Making the mean field approximation to the PNJL Lagrangian 
and performing the path integration of 
the resultant partition function over $q$ under 
the twisted boundary condition~(\ref{eq:BC}), 
one can obtain the thermodynamical potential 
\begin{align}
\Omega
=& -2 N_{\rm f} \int_\Lambda \frac{d^3 p}{(2\pi)^3}
   \Bigl[ 3 E_p \notag \\
&+ \frac{1}{\beta}
           \ln~ [1 + 3(\Phi+{\bar \Phi} e^{-\beta E^-_{\bf p}}) 
           e^{-\beta E^-_{\bf p}}+ e^{-3\beta E^-_{\bf p}}] \notag\\
&+ \frac{1}{\beta} 
           \ln~ [1 + 3({\bar \Phi}+{\Phi e^{-\beta E^+_{\bf p}}}) 
              e^{-\beta E^+_{\bf p}}+ e^{-3\beta E^+_{\bf p}}]
	      \Bigl]
	      \nonumber\\
&+G_{\rm s} \sigma^2-G_{\rm v} \omega^2+3G_{\rm s8} \sigma^4
+{\cal U}, 
\label{PNJL-Omega}
\end{align}
where $\sigma = \langle {\bar q} q \rangle$, 
$\omega = \langle {\bar q} \gamma_0 q \rangle$ and 
$E^\pm_{\bf p}=E_{\bf p} \pm iT\theta_\varphi$ with 
$E_{\bf p}=\sqrt{{\bf p}^2+M^2}$, 
$M=m_0-2G_{\rm s}\sigma-4G_\mathrm{s8}\sigma^3$ and
$iT\theta_{\varphi}=- 2G_{\rm v}\omega - i\pi T + iT\varphi$. 
Here, $m_0$ is the current quark mass and we use $m_0=5.5$ MeV.
The constants $G_{\rm s}$, $G_{\rm v}$, $G_{\rm s8}$ 
denote coupling strengths of the scalar-type four-quark, 
the vector-type four-quark 
and the scalar-type eight-quark interaction, respectively.
The Polyakov-loop potential ${\cal U}$ is a function of $\Phi$ and 
its Hermitian conjugate ${\bar \Phi}$.  
The 3-dimensional momentum integration is regularized by a cutoff $\Lambda$. 
The classical variables $X=\sigma, \omega, \Phi$, ${\bar \Phi}$ are 
determined by solving the stationary conditions 
$\partial \Omega / \partial X =0$ numerically, and $\Sigma^{(n)}$ is 
obtained numerically from $\sigma(\varphi)$ with \eqref{DPL}. 
Details of the PNJL model are shown in Ref.~\cite{Kashiwa1}.

In the PNJL calculations, two types of ${\cal U}$ are often used; 
one has the polynomial form~\cite{Ratti} and the other has
the logarithm form~\cite{Rossner}. 
The former (latter) potential is referred to as RTW05 (RRW06). 
Parameters of these potentials are fitted to LQCD data at finite $T$ 
in the pure gauge limit~\cite{Boyd1,Kaczmarek1}, but 
one of the parameters, $T_0$, is usually refitted to 
reproduce the pseudo-critical temperature $T_{\rm c}=173\pm8$~MeV 
in full LQCD~\cite{Karsch}. 
The value of $T_0$ thus determined is $185$ ($200$ MeV) 
for the RTW05 (RRW06) potential. 
The RRW06 potential can reproduce LQCD data~\cite{FP,Elia,Chen} at imaginary 
chemical potential, but the RTW05 does not~\cite{Sakai2}. 
Therefore, the RRW06 potential is mainly used in the present PNJL analyses.

We do four types of PNJL calculations shown in Table~\ref{Table-S}.
In PNJL-I, 
$G_{\rm s}$ and $\Lambda$ are adjusted to the pion mass $M_\pi=138$ MeV 
and the pion decay constant and $f_\pi=93.3$ MeV at $T=0$; see 
Ref.~\cite{Kashiwa1} for values of the parameters. 
In PNJL-II, -III and -IV, 
$G_{\rm s}$, $G_{\rm s8}$, $\Lambda$ are fitted to 
$M_\pi=138$ MeV and $f_\pi=93.3$ MeV and the sigma meson mass 
$M_\sigma=600$ MeV; see Ref.~\cite{Kashiwa1} for values of the parameters. 
Since the empirical value of $M_\sigma$ has a large error bar, the empirical value is not a strong constraint to determine 
the strength $G_{\rm s8}$. 
We then think $G_{\rm s8}$ 
as a nearly-free parameter. The value of $G_{\rm s8}$ determined above is 
just an example.  
In PNJL-III, 
the vector-type interaction is a free parameter. 
The strength of the vector-type interaction was determined to reproduce 
LQCD data~\cite{FP,Elia,Chen} at imaginary chemical potential~\cite{Sakai2}. 
The ratio $G_{\rm v}/G_{\rm s}$ is 0.85. As an example, 
we take a bit smaller value of $G_{\rm v}/G_{\rm s}=0.667$. 
For comparison, we also do the NJL calculation 
with the scalar-type eight-quark interaction; this is referred to as NJL. 
Since NJL agrees with PNJL-II in the limit of $T=0$~\cite{Sakai},  
the same parameter set is taken in the two models.

\begin{table}[h]
\begin{center}
\begin{tabular}{ccccccc}
\hline
\hline
Model & Interaction & ${\cal U}$ & $T_\mathrm{c}^\Phi$ & $T_\mathrm{c}^\sigma$ & $T_\mathrm{c}^{\Sigma^{(1)}}$ & line  \\ 
\hline
PNJL-I~~ & $\sigma^2$ & RRW06 & $1.01T_\mathrm{c}$ & $1.32T_\mathrm{c}$ & $1.01T_\mathrm{c}$ & thin-solid\\ 
\hline
PNJL-II~ & $\sigma^2,~\sigma^4$ & RRW06 & $T_\mathrm{c}$ & $1.14T_\mathrm{c}$ & $T_\mathrm{c}$ & solid\\ 
\hline
PNJL-III & ~$\sigma^2,~\sigma^4,~\omega^2$~ & RRW06 & $T_\mathrm{c}$ & $1.14T_\mathrm{c}$ & $T_\mathrm{c}$ & dashed\\ 
\hline
PNJL-IV~ & $\sigma^2,~\sigma^4$ & RTW05 & $0.98T_\mathrm{c}$ & $1.10T_\mathrm{c}$ & $1.11T_\mathrm{c}$ & dotted\\ 
\hline
NJL & $\sigma^2,~\sigma^4$ & - & - & $0.94T_\mathrm{c}$ & $0.97T_\mathrm{c}$ & dot-dashed\\ 
\hline
\hline
\end{tabular}
\caption
{
Definitions and results of PNJL and NJL calculations. 
Symbols, RTW05 and RRW06, are the Polyakov potentials 
of polynomial type~\cite{Ratti} and logarithm type~\cite{Rossner}, 
respectively,
while $\sigma^2$, $\sigma^4$, 
$\omega^2$ denote the scalar-type four-quark, the scalar-type eight-quark and the vector-type four-quark interaction, respectively. 
Symbols $T_\mathrm{c}^\sigma$, $T_\mathrm{c}^\Phi$, 
$T_\mathrm{c}^{\Sigma^{(1)}}$ denote the pseudo-critical temperatures 
defined by peak positions of $d\sigma/dT$, $d\Phi/dT$, $d\Sigma^{(1)}/dT$, 
respectively. Here, $T_\mathrm{c}=173$ MeV. 
\label{Table-S}}
\end{center}
\end{table}


As mentioned above, the twisted boundary condition 
\eqref{eq:BC} is imposed on $D_{\varphi}$. 
Therefore, $\Phi$ and ${\bar \Phi}$ are first obtained 
under the anti-periodic boundary condition $\varphi=\pi$. 
The quantities $\sigma(\varphi)$ and $\omega(\varphi)$ 
are determined by solving the stationary conditions, 
$\partial \Omega/\partial \sigma=0$ and $\partial \Omega/\partial \omega=0$, 
numerically under the twisted boundary condition \eqref{eq:BC} with 
$\Phi$ and ${\bar \Phi}$ fixed to the values determined 
at $\varphi=\pi$.

The pseudo-critical temperature $T_{\rm c}^\Phi$ of the deconfinement 
crossover is usually defined by a peak position of 
either the Polyakov-loop susceptibility or $d\Phi/dT$. In the present 
analysis, we take the latter to compare $T_{\rm c}^\Phi$ with 
the pseudo-critical temperature $T_{\rm c}^{\Sigma^{(1)}}$ estimated by 
a peak position of ${d\Sigma^{(1)}/{dT}}$, 
because we can not define the susceptibility for the dressed Polyakov loop 
$\Sigma^{(1)}$. 
For consistency,
the pseudo-critical temperature $T_{\rm c}^\sigma$ of 
the chiral crossover is defined by a peak position 
of $d\sigma/dT$. 
In the PNJL calculations with the RRW06 potential, 
$d\sigma/dT$ has two peaks, but 
we take the second peak close to a peak position of the chiral susceptibility.

Table~\ref{Table-S} summarizes values 
of $T_{\rm c}^\Phi$, $T_{\rm c}^\sigma$, $T_{\rm c}^{\Sigma^{(1)}}$ in 
five types of model calculations, 
where the values are normalized by $T_{\rm c}=173$~MeV, i.e. 
the LQCD result~\cite{Karsch} for $T_{\rm c}^\Phi$ and $T_{\rm c}^\sigma$. 
In PNJL-I, -II, -III with the RRW06 potential, 
$T_{\rm c}^\Phi$ and $T_{\rm c}^{\Sigma^{(1)}}$ are close to each other 
and also to the LQCD data~\cite{Karsch}, 
although this property is not seen in PNJL-IV with the RTW05 potential. 
As for $T_{\rm c}^\sigma$, PNJL-II is more consistent 
with the LQCD data than PNJL-I and hence 
the scalar-type eight-quark interaction is necessary 
to reproduce the LQCD data.

\begin{figure}[htbp]
\begin{center}
\includegraphics[width=0.38\textwidth]{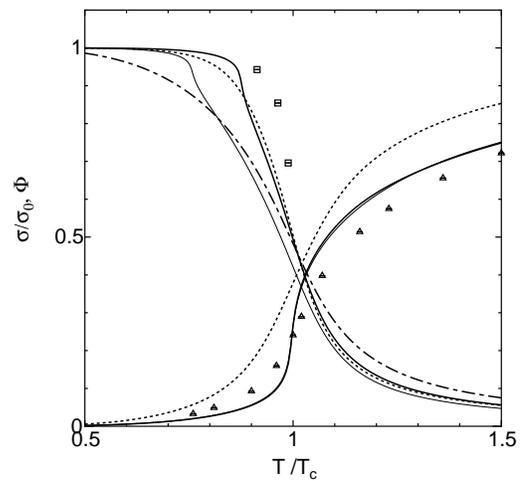}
\end{center}
\caption{ Chiral condensate as a function of $T/T_{\rm c}^{\sigma}$ 
and renormalized Polyakov loop as a function of $T/T_{\rm c}^{\Phi}$ in 
the case of $\varphi=\pi$. 
Here, $\sigma$ is normalized by the value $\sigma_0$ 
at $T=0$. The values of 
$T_{\rm c}^{\sigma}$ and $T_{\rm c}^{\Phi}$ are
summarized in Table.~\ref{Table-S}.
Definitions of lines are also shown in Table.~\ref{Table-S}.
LQCD data shown by box and triangle symbols are taken from 
Ref.~\cite{Boyd2} and Ref.~\cite{Kaczmarek2}, respectively.
}
\label{Fig:MU0}
\end{figure}

Figure~\ref{Fig:MU0} presents $T$-dependence of the chiral condensate 
and that of renormalized $\Phi$, 
where the anti-periodic boundary condition $\varphi=\pi$ is taken. 
The resultant chiral condensate $\sigma$ is 
normalized by the value $\sigma_0$ at $T=0$. 
Temperature $T$ is normalized by 
$T_{\rm c}^{\sigma}$ for the chiral condensate and by $T_{\rm c}^{\Phi}$ for renormalized $\Phi$ 
to compare the shapes of 
$\sigma/\sigma_0$ and $\Phi$ with LQCD data~\cite{Boyd2,Kaczmarek2}. 
Also for the shape for $\sigma$, PNJL-II (solid curve) 
gives a better agreement with the LQCD data than PNJL-I (thin solid curve). 
Meanwhile, the solid and thin solid curves are 
almost identical for $\Phi$, indicating that 
the scalar-type eight-quark interaction hardly affects the shape of $\Phi$. 
At zero quark chemical potential, the vector-type four-quark interaction 
does not affect $\sigma$ and $\Phi$ in the mean field approximation; hence, 
PNJL-III yields the same $\sigma$ and $\Phi$ as PNJL-II in Fig.~\ref{Fig:MU0}. 
Nevertheless, the interaction gives a sizable effect on $\Sigma^{(1)}$, 
as shown later. 
For the shape of $\Phi$, 
PNJL-IV (dotted curve) gives 
a larger disagreement with the LQCD data than PNJL-II. Thus, 
the RRW06 potential is better than the RTW05 potential also for the shape of 
$\Phi$.

Full LQCD simulation of Ref.~\cite{Bilgici3} shows that 
$T_{\rm c}^{\Sigma^{(1)}} \simeq T_{\rm c}^{\sigma} \simeq 
T_{\rm c}^{\Phi} \simeq 153$~MeV, 
while that of Ref.~\cite{Karsch} does 
$T_{\rm c}^{\sigma} \simeq T_{\rm c}^{\Phi} \simeq 173$~MeV. 
The two LQCD results have a systematic error of $\sim 10~\%$. 
Moreover, it is reported in Ref.~\cite{Aoki} 
for the $2+1$-flavor system 
that there exists a non-negligible deviation between 
$T_{\rm c}^{\sigma}$ and $T_{\rm c}^{\Phi}$. This indicates that it 
is an unsettled problem 
whether $T_{\rm c}^{\sigma}$ and $T_{\rm c}^{\Phi}$ really coincide or not. 
In this work, however, we assume the coincidence, 
since the LQCD data \cite{Bilgici3} that we are analyzing 
has the property. 
In the present analysis, temperature is normalized by 
$T_{\rm c}$, but it is taken to be 153~MeV for LQCD data 
of Ref.~\cite{Bilgici3} and 173~MeV for model calculations and LQCD data of 
Ref.~\cite{Karsch,Boyd2,Kaczmarek2}.

\begin{figure}[htbp]
\begin{center}
 \includegraphics[width=0.38\textwidth]{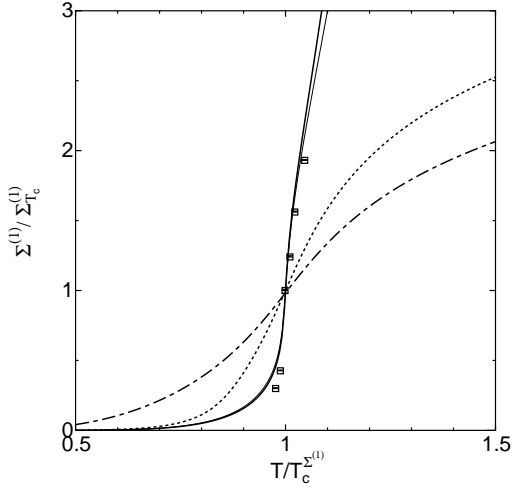}  
\end{center}
\caption{Normalized dressed Polyakov loop as a function of 
$T/T_{\rm c}^{\Sigma^{(1)}}$. 
Definitions of lines are shown in Table.~\ref{Table-S}.
LQCD data (box symbols) are taken from Ref.~\cite{Bilgici3}.
}
\label{Fig:D-CC}
\end{figure}

Figure~\ref{Fig:D-CC} presents the normalized 
dressed Polyakov loop $\Sigma^{(1)}/\Sigma^{(1)}_{T_{\rm c}}$ 
as a function of $T/T_{\rm c}^{\Sigma^{(1)}}$, 
where $\Sigma^{(1)}_{T_{\rm c}}$ is the dressed Polyakov loop at 
$T=T_{\rm c}^{\Sigma^{(1)}}$. 
PNJL-I (thin solid line) and 
PNJL-II (solid line) well reproduces 
LQCD data~\cite{Bilgici3} (box symbols), 
but PNJL-IV (dotted line) and 
NJL (dot-dashed line) do not. 
Near $T_{\rm c}^{\Sigma^{(1)}}$, 
the scalar-type eight-quark interaction hardly affects 
the dressed Polyakov loop. 
Thus, the RRW06 potential is necessary to explain $T$ dependence of 
$\Phi$, $\sigma$ and $\Sigma^{(1)}$ consistently.

\begin{figure}[htbp]
\begin{center}
 \includegraphics[width=0.38\textwidth]{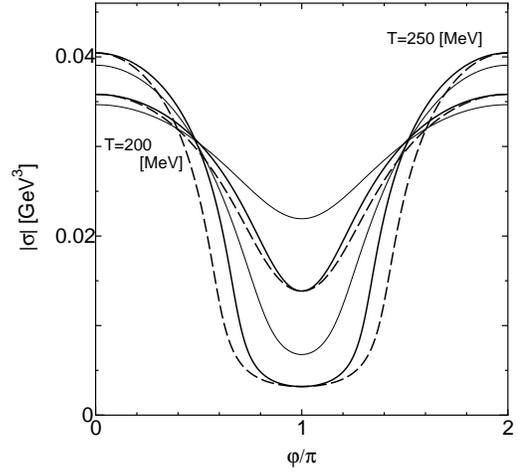} 
\end{center}
\caption{
$\varphi$-dependence of the chiral condensate at 
$T=200$ and 250~MeV. 
Definitions of lines are shown in Table.~\ref{Table-S}.
}
\label{Fig:DQC-SV-1}
\end{figure}

\begin{figure}[htbp]
\begin{center}
 \includegraphics[width=0.38\textwidth]{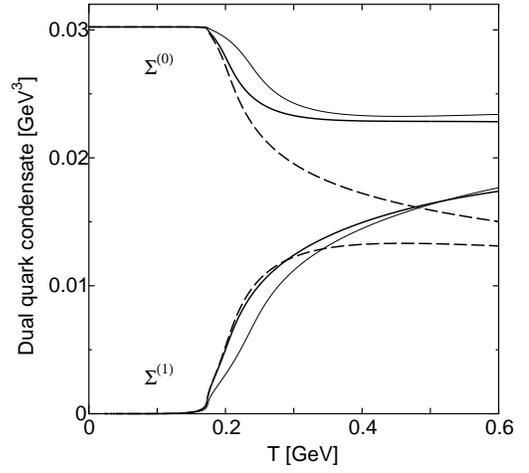} 
\end{center}
\caption{
$T$-dependence of the dual quark condensate. 
Definitions of lines are shown in Table.~\ref{Table-S}.
}
\label{Fig:DQC-SV-2}
\end{figure}

Figure~\ref{Fig:DQC-SV-1} shows $\varphi$-dependence 
of $\sigma$ at $T=200$ and $250$ MeV, while Fig.~\ref{Fig:DQC-SV-2} represented
$T$-dependence of the dual quark condensates with $n=0$ and $1$. 
The difference between PNJL-I (thin-solid line) and 
PNJL-II (solid line) shows an effect of 
the scalar-type eight-quark interaction. 
As shown in Fig.~\ref{Fig:DQC-SV-1}, 
the effect on $\sigma$ is large around $\varphi=\pi$, 
but small around $\varphi=0$. 
The absolute value of $\sigma$ is relatively larger 
around $\varphi=0$. As a consequence, as shown in Fig.~\ref{Fig:DQC-SV-2}, 
the effect on $\Sigma^{(0)}$ and 
$\Sigma^{(1)}$ are appreciable only 
at $T_{\rm c}^{\Sigma^{(1)}}<T<2T_{\rm c}^{\Sigma^{(1)}}$. Thus, 
the effect of the eight-quark interaction becomes more appreciable for 
$\sigma$ at $\varphi=\pi$ than for $\Sigma^{(0)}$ and $\Sigma^{(1)}$. 
At $T>2T_{\rm c}^{\Sigma^{(1)}}$, the effect becomes negligible, 
since $\sigma$ itself is tiny there. 
Meanwhile, the difference between PNJL-II (solid line) and PNJL-III 
(dashed line) presents an effect of the vector-type four-quark interaction. 
The effect on $\sigma$ is zero at $\varphi=\pi$, and appreciable at nonzero $\varphi(\neq \pi)$, as shown in Fig.~\ref{Fig:DQC-SV-1}. 
The effect on $\Sigma^{(0)}$ and $\Sigma^{(1)}$ is then appreciable at $T_{\rm c}^{\Sigma^{(1)}}<T<1.5T_{\rm c}^{\Sigma^{(1)}}$ and sizable at $T>1.5T_{\rm c}^{\Sigma^{(1)}}$, as shown in Fig.~\ref{Fig:DQC-SV-2}.
Thus, the effect of the vector-type four-quark interaction becomes more appreciable for $\Sigma^{(0)}$ and $\Sigma^{(1)}$ than for $\sigma$ at $\varphi=\pi$. 
This indicates that the dual quark condensate 
at $T>1.5T_{\rm c}^{\Sigma^{(1)}}$ is a good quantity 
to determine the strength of the vector-type four-quark interaction.

The present PNJL results are consistent with the LQCD data 
for $\Sigma^{(1)}/\Sigma^{(1)}_{T_{\rm c}}$ near $T_{\rm c}$, 
but not for $\sigma(\varphi)$ itself even if both are compared at the same values of $T/T^{\Sigma^{(1)}}_c$, 
although the latter is not shown explicitly in this paper. 
In LQCD, $\sigma(\varphi)$ and $\Sigma^{(1)}$ are not renormalized, 
while $\Phi$ presented in Fig.~\ref{Fig:MU0} is renormalized. 
The reasonable agreement of the PNJL result with 
the LQCD one for $\sigma/\sigma_0$ and $\Phi$ 
in Fig.~\ref{Fig:MU0} and $\Sigma^{(1)}/\Sigma^{(1)}_{T_{\rm c}}$ 
in Fig.~\ref{Fig:D-CC} may imply that 
$\sigma/\sigma_0$ and $\Sigma^{(1)}/\Sigma^{(1)}_{T_{\rm c}}$ have 
better renormalization properties than $\sigma(\varphi)$ itself.

Finally, we discuss the sensitivity of $\Sigma^{(1)}$ to the parameters of 
the PNJL model. 
If $\Lambda$ is varied by 10 \%, the pion decay constant at $T=0$ 
is changed by about 20~\%. Hence, one can change 
$\Lambda$ by only 0.1~\% in order that 
the calculated pion decay constant is consistent with the observed one with 
0.1~\% error. 
This is also the case for $G_{\rm s}$. 
If $\Lambda$ and $G_{\rm s}$ are varied by 1 \%, $\Sigma^{(1)}$ are changed by 
about 10~\%, but the normalized quantity 
$\Sigma^{(1)}/\Sigma^{(1)}_{T_{\rm c}}$ is hardly changed 
near $T_{\rm c}$. Further, $\Sigma^{(1)}$ near $T_{\rm c}$ is much less 
sensitive to $G_{\rm s8}$ and $G_{\rm v}$, although these are 
nearly-free parameters. Thus, we can think that 
$\Sigma^{(1)}/\Sigma^{(1)}_{T_{\rm c}}$ near $T_{\rm c}$ changes little
from the present values.

K.K. is supported by the Japan Society for the Promotion of Science for Young Scientists.



\begin{thebibliography}{19}
\expandafter\ifx\csname natexlab\endcsname\relax\def\natexlab#1{#1}\fi
\expandafter\ifx\csname bibnamefont\endcsname\relax
  \def\bibnamefont#1{#1}\fi
\expandafter\ifx\csname bibfnamefont\endcsname\relax
  \def\bibfnamefont#1{#1}\fi
\expandafter\ifx\csname citenamefont\endcsname\relax
  \def\citenamefont#1{#1}\fi
\expandafter\ifx\csname url\endcsname\relax
  \def\url#1{\texttt{#1}}\fi
\expandafter\ifx\csname urlprefix\endcsname\relax\def\urlprefix{URL }\fi
\providecommand{\bibinfo}[2]{#2}
\providecommand{\eprint}[2][]{\url{#2}}


\bibitem[{\citenamefont{Bilgici et al.}(2008)}]{Bilgici1}
\bibinfo{author}{\bibfnamefont{E.}~\bibnamefont{Bilgici}},
\bibinfo{author}{\bibfnamefont{F.}~\bibnamefont{Bruckmann}},
\bibinfo{author}{\bibfnamefont{C.}~\bibnamefont{Gattringer}},
\bibnamefont{and}
\bibinfo{author}{\bibfnamefont{C.}~\bibnamefont{Hagen}},  
  \bibinfo{journal}{Phys. Rev.\ D} \textbf{\bibinfo{volume}{77}},
  \bibinfo{pages}{094007} (\bibinfo{year}{2008}). 

\bibitem[{\citenamefont{Bilgici et al.}(2008)}]{Bilgici2}
\bibinfo{author}{\bibfnamefont{F.}~\bibnamefont{Bruckmann}},
\bibinfo{author}{\bibfnamefont{C.}~\bibnamefont{Hagen}},  
\bibinfo{author}{\bibfnamefont{E.}~\bibnamefont{Bilgici}},
\bibnamefont{and}
\bibinfo{author}{\bibfnamefont{C.}~\bibnamefont{Gattringer}},
  \bibinfo{journal}{Pos} \textbf{\bibinfo{volume}{Lattice2008}},
  \bibinfo{pages}{262} (\bibinfo{year}{2008});
  \bibinfo{journal}{Pos} \textbf{\bibinfo{volume}{Confinement8}},
  \bibinfo{pages}{054} (\bibinfo{year}{2008}). 
%
\bibitem[{\citenamefont{Bilgici et al.}(2009)}]{Bilgici3}
\bibinfo{author}{\bibfnamefont{E.}~\bibnamefont{Bilgici}},  
\bibinfo{author}{\bibfnamefont{F.}~\bibnamefont{Bruckmann}},
\bibinfo{author}{\bibfnamefont{J.}~\bibnamefont{Danzer}},
\bibinfo{author}{\bibfnamefont{C.}~\bibnamefont{Gattringer}},
\bibinfo{author}{\bibfnamefont{C.}~\bibnamefont{Hagen}},
\bibinfo{author}{\bibfnamefont{E.}~\bibfnamefont{M.}~\bibnamefont{Ilgenfritz}},
\bibnamefont{and}
\bibinfo{author}{\bibfnamefont{A.}~\bibnamefont{Maas}},
\bibinfo{howpublished}{arXiv:0906.3957};
\bibinfo{author}{\bibfnamefont{E.}~\bibnamefont{Bilgici}},  
\bibinfo{howpublished}{PhD Thesis, University of Graz, Austria, 2009
(http://physik.uni-graz.at/itp/files/bilgici/dissertation.pdf).}
%
\bibitem[{\citenamefont{Fukushima}(2004)}]{Fukushima1}
\bibinfo{author}{\bibfnamefont{K.}~\bibnamefont{Fukushima}}, 
  \bibinfo{journal}{Phys. Lett.\ B} \textbf{\bibinfo{volume}{591}},
  \bibinfo{pages}{277} (\bibinfo{year}{2004});
   \bibinfo{journal}{Phys. Rev. D} \textbf{\bibinfo{volume}{77}},
  \bibinfo{pages}{114028} (\bibinfo{year}{2008});
  \bibinfo{journal}{Phys. Rev. D} \textbf{\bibinfo{volume}{78}},
  \bibinfo{pages}{114019} (\bibinfo{year}{2008}).
%
\bibitem[{\citenamefont{Ratti et al.}(2006)}]{Ratti}
\bibinfo{author}{\bibfnamefont{C.}~\bibnamefont{Ratti}},
\bibinfo{author}{\bibfnamefont{M.}~\bibfnamefont{A.}~\bibnamefont{Thaler}},
\bibnamefont{and}
\bibinfo{author}{\bibfnamefont{W.}~\bibnamefont{Weise}},  
  \bibinfo{journal}{Phys. Rev.\ D} \textbf{\bibinfo{volume}{73}},
  \bibinfo{pages}{014019} (\bibinfo{year}{2006}). 

\bibitem[{\citenamefont{Rossner et al.}(2007)}]{Rossner}
\bibinfo{author}{\bibfnamefont{S.}~\bibnamefont{R\"{o}{\ss}ner}},
\bibinfo{author}{\bibfnamefont{C.}~\bibnamefont{Ratti}},
\bibnamefont{and}
\bibinfo{author}{\bibfnamefont{W.}~\bibnamefont{Weise}},  
  \bibinfo{journal}{Phys. Rev.\ D} \textbf{\bibinfo{volume}{75}},
  \bibinfo{pages}{034007} (\bibinfo{year}{2007}). 
%
\bibitem[{\citenamefont{Costa et al}(2008)}]{Costa}
\bibinfo{author}{\bibfnamefont{P.}~\bibnamefont{Costa}}, 
\bibinfo{author}{\bibfnamefont{M.}~\bibfnamefont{C.}~\bibnamefont{Ruivo}}, 
\bibinfo{author}{\bibfnamefont{C.}~\bibfnamefont{A.}~\bibfnamefont{de}~\bibnamefont{Sousa}}, 
\bibinfo{author}{\bibfnamefont{H.}~\bibnamefont{Hansen}},
\bibnamefont{and}
\bibinfo{author}{\bibfnamefont{W.}~\bibfnamefont{M.}~\bibnamefont{Alberico}},
  \bibinfo{journal}{Phys. Rev.\ D} \textbf{\bibinfo{volume}{79}},
  \bibinfo{pages}{116003} (\bibinfo{year}{2009}). 
%
\bibitem[{\citenamefont{Kashiwa et al}(2008)}]{Kashiwa1}
\bibinfo{author}{\bibfnamefont{K.}~\bibnamefont{Kashiwa}}, 
\bibinfo{author}{\bibfnamefont{H.}~\bibnamefont{Kouno}}, 
\bibinfo{author}{\bibfnamefont{M.}~\bibnamefont{Matsuzaki}}, 
\bibnamefont{and}
\bibinfo{author}{\bibfnamefont{M.}~\bibnamefont{Yahiro}},
  \bibinfo{journal}{Phys.\ Lett.\ B} \textbf{\bibinfo{volume}{662}},
  \bibinfo{pages}{26} (\bibinfo{year}{2008});
  \bibinfo{author}{\bibfnamefont{K.}~\bibnamefont{Kashiwa}}, 
\bibinfo{author}{\bibfnamefont{M.}~\bibnamefont{Matsuzaki}}, 
\bibinfo{author}{\bibfnamefont{H.}~\bibnamefont{Kouno}}, 
\bibinfo{author}{\bibfnamefont{Y.}~\bibnamefont{Sakai}},
\bibnamefont{and}
\bibinfo{author}{\bibfnamefont{M.}~\bibnamefont{Yahiro}},
  \bibinfo{journal}{Phys.\ Rev.\  D} \textbf{\bibinfo{volume}{79}},
  \bibinfo{pages}{076008} (\bibinfo{year}{2009}).
%
\bibitem[{\citenamefont{Sakai et al}(2008)}]{Sakai}
\bibinfo{author}{\bibfnamefont{Y.}~\bibnamefont{Sakai}},
\bibinfo{author}{\bibfnamefont{K.}~\bibnamefont{Kashiwa}}, 
\bibinfo{author}{\bibfnamefont{H.}~\bibnamefont{Kouno}}, 
\bibnamefont{and}
\bibinfo{author}{\bibfnamefont{M.}~\bibnamefont{Yahiro}},
  \bibinfo{journal}{Phys.\ Rev.\  D} \textbf{\bibinfo{volume}{77}},
  \bibinfo{pages}{051901(R)} (\bibinfo{year}{2008});
%
  \bibinfo{journal}{Phys.\ Rev.\  D} \textbf{\bibinfo{volume}{78}},
  \bibinfo{pages}{ 036001} (\bibinfo{year}{2008});
%
\bibinfo{author}{\bibfnamefont{Y.}~\bibnamefont{Sakai}},
\bibinfo{author}{\bibfnamefont{K.}~\bibnamefont{Kashiwa}}, 
\bibinfo{author}{\bibfnamefont{H.}~\bibnamefont{Kouno}}, 
\bibinfo{author}{\bibfnamefont{M.}~\bibnamefont{Matsuzaki}}, 
\bibnamefont{and}
\bibinfo{author}{\bibfnamefont{M.}~\bibnamefont{Yahiro}},
  \bibinfo{journal}{Phys.\ Rev.\  D} \textbf{\bibinfo{volume}{78}},
  \bibinfo{pages}{076007} (\bibinfo{year}{2008}).
%
\bibitem[{\citenamefont{Sakai et al}(2008)}]{Sakai2}
\bibinfo{author}{\bibfnamefont{Y.}~\bibnamefont{Sakai}},
\bibinfo{author}{\bibfnamefont{K.}~\bibnamefont{Kashiwa}}, 
\bibinfo{author}{\bibfnamefont{H.}~\bibnamefont{Kouno}}, 
\bibinfo{author}{\bibfnamefont{M.}~\bibnamefont{Matsuzaki}}, 
\bibnamefont{and}
\bibinfo{author}{\bibfnamefont{M.}~\bibnamefont{Yahiro}},
  \bibinfo{journal}{Phys.\ Rev.\  D} \textbf{\bibinfo{volume}{79}},
  \bibinfo{pages}{096001} (\bibinfo{year}{2009}).
%
\bibitem[{\citenamefont{Abuki}(2008)}]{Abuki}
\bibinfo{author}{\bibfnamefont{H.}~\bibnamefont{Abuki}},
\bibinfo{author}{\bibfnamefont{R.}~\bibnamefont{Anglani}},
\bibinfo{author}{\bibfnamefont{R.}~\bibnamefont{Gatto}},
\bibinfo{author}{\bibfnamefont{G.}~\bibnamefont{Nardulli}},
\bibnamefont{and}
\bibinfo{author}{\bibfnamefont{M.}~\bibnamefont{Ruggieri}},
 \bibinfo{journal}{Phys.\ Rev.\  D} \textbf{\bibinfo{volume}{78}},
  \bibinfo{pages}{034034} (\bibinfo{year}{2008});
\bibinfo{author}{\bibfnamefont{H.}~\bibnamefont{Abuki}},
\bibinfo{author}{\bibfnamefont{M.}~\bibnamefont{Ciminale}},
\bibinfo{author}{\bibfnamefont{R.}~\bibnamefont{Gatto}},
\bibnamefont{and}
\bibinfo{author}{\bibfnamefont{M.}~\bibnamefont{Ruggieri}},
  \bibinfo{journal}{Phys.\ Rev.\  D} \textbf{\bibinfo{volume}{79}},
  \bibinfo{pages}{034021} (\bibinfo{year}{2009}).
%
\bibitem[{\citenamefont{Hell et al.}(2009)}]{Hell}
\bibinfo{author}{\bibfnamefont{T.}~\bibnamefont{Hell}},
\bibinfo{author}{\bibfnamefont{S.}~\bibnamefont{R\"{o}{\ss}ner}},
\bibinfo{author}{\bibfnamefont{M.}~\bibnamefont{Cristoforetti}},
\bibnamefont{and}
\bibinfo{author}{\bibfnamefont{W.}~\bibnamefont{Weise}},
  \bibinfo{journal}{Phys. Rev. D} \textbf{\bibinfo{volume}{79}},
  \bibinfo{pages}{014022} (\bibinfo{year}{2009}). 
%
\bibitem[{\citenamefont{Kashiwa et al}(2009)}]{Kashiwa2}
\bibinfo{author}{\bibfnamefont{K.}~\bibnamefont{Kashiwa}},
\bibinfo{author}{\bibfnamefont{M.}~\bibnamefont{Yahiro}},
\bibinfo{author}{\bibfnamefont{H.}~\bibnamefont{Kouno}},
\bibinfo{author}{\bibfnamefont{M.}~\bibnamefont{Matsuzaki}}, 
\bibnamefont{and}
\bibinfo{author}{\bibfnamefont{Y.}~\bibnamefont{Sakai}},
   \bibinfo{journal}{J. Phys. G} \textbf{\bibinfo{volume}{36}},
  \bibinfo{pages}{105001} (\bibinfo{year}{2009}).
%
\bibitem[{\citenamefont{Kouno et al}(2009)}]{Kouno}
\bibinfo{author}{\bibfnamefont{H.}~\bibnamefont{Kouno}},
\bibinfo{author}{\bibfnamefont{Y.}~\bibnamefont{Sakai}},
\bibinfo{author}{\bibfnamefont{K.}~\bibnamefont{Kashiwa}},
\bibnamefont{and}
\bibinfo{author}{\bibfnamefont{M.}~\bibnamefont{Yahiro}},
   \bibinfo{journal}{J. Phys. G} \textbf{\bibinfo{volume}{36}},
  \bibinfo{pages}{115010} (\bibinfo{year}{2009}).
%
\bibitem[{\citenamefont{Nam}(2009)}]{Nam}
\bibinfo{author}{\bibfnamefont{S.}~\bibnamefont{i.}~\bibnamefont{Nam}}, 
 \bibinfo{howpublished}{arXiv:hep-ph/0905.3609}.
%
\bibitem[{\citenamefont{Fischer}(2009)}]{Fischer}
\bibinfo{author}{\bibfnamefont{C.}~\bibnamefont{S}~\bibnamefont{Fischer}}, 
 \bibinfo{journal}{Phys.\ Rev.\  Lett} \textbf{\bibinfo{volume}{103}},
 \bibinfo{pages}{052003} (\bibinfo{year}{2009});
%
\bibinfo{author}{\bibfnamefont{C.}~\bibnamefont{S}~\bibnamefont{Fischer}},
\bibnamefont{and}
\bibinfo{author}{\bibfnamefont{J.}~\bibnamefont{A.}~\bibnamefont{Mueller}},
  \bibinfo{journal}{Phys.\ Rev.\  D} \textbf{\bibinfo{volume}{80}},
  \bibinfo{pages}{074029} (\bibinfo{year}{2009}).
%
\bibitem[{\citenamefont{Braun}(2009)}]{Braun}
\bibinfo{author}{\bibfnamefont{J.}~\bibnamefont{Braun}}, 
\bibinfo{author}{\bibfnamefont{L.}~\bibnamefont{M.}~\bibnamefont{Haas}},
\bibinfo{author}{\bibfnamefont{F.}~\bibnamefont{Marhauser}},
\bibnamefont{and}
\bibinfo{author}{\bibfnamefont{J.}~\bibnamefont{M.}~\bibnamefont{Pawlowski}},
 \bibinfo{howpublished}{arXiv:hep-ph/0908.0008}.
%
\bibitem[{\citenamefont{Synatschke et al}(2008)}]{Synatschke}
\bibinfo{author}{\bibfnamefont{F.}~\bibnamefont{Synatschke}},
\bibinfo{author}{\bibfnamefont{A.}~\bibnamefont{Wipf}}, 
\bibnamefont{and}
\bibinfo{author}{\bibfnamefont{K.}~\bibnamefont{Langfeld}},
  \bibinfo{journal}{Phys.\ Rev.\  D} \textbf{\bibinfo{volume}{77}},
  \bibinfo{pages}{114018} (\bibinfo{year}{2008}).
%
\bibitem[{\citenamefont{Kashiwa et al}(2007)}]{Kashiwa3}
\bibinfo{author}{\bibfnamefont{K.}~\bibnamefont{Kashiwa}}, 
\bibinfo{author}{\bibfnamefont{H.}~\bibnamefont{Kouno}}, 
\bibinfo{author}{\bibfnamefont{T.}~\bibnamefont{Sakaguchi}}, 
\bibinfo{author}{\bibfnamefont{M.}~\bibnamefont{Matsuzaki}}, 
\bibnamefont{and}
\bibinfo{author}{\bibfnamefont{M.}~\bibnamefont{Yahiro}},
\bibinfo{journal}{Phys. Lett.\ B} \textbf{\bibinfo{volume}{647}},
\bibinfo{pages}{446} (\bibinfo{year}{2007}).
%
\bibitem[{\citenamefont{Forcrand and Philipsen}(2002)}]{FP}
\bibinfo{author}{\bibfnamefont{P.}~\bibnamefont{de}~\bibnamefont{Forcrand}} 
\bibnamefont{and}
\bibinfo{author}{\bibfnamefont{O.}~\bibnamefont{Philipsen}},  
\bibinfo{journal}{Nucl. Phys. } \textbf{\bibinfo{volume}{B642}},
\bibinfo{pages}{290} (\bibinfo{year}{2002});
\bibinfo{journal}{Nucl. Phys. } \textbf{\bibinfo{volume}{B673}},
\bibinfo{pages}{170} (\bibinfo{year}{2003}). 
%
\bibitem[{\citenamefont{Elia and Lombardo}(2003)}]{Elia}
\bibinfo{author}{\bibfnamefont{M.}~\bibnamefont{D'Elia}} \bibnamefont{and}
\bibinfo{author}{\bibfnamefont{M.}~\bibfnamefont{P.}~\bibnamefont{Lombardo}},  
\bibinfo{journal}{Phys. Rev.\  D} \textbf{\bibinfo{volume}{67}},
\bibinfo{pages}{014505} (\bibinfo{year}{2003}); 
\bibinfo{journal}{Phys. Rev.\ D} \textbf{\bibinfo{volume}{70}},
\bibinfo{pages}{074509} (\bibinfo{year}{2004});
\bibinfo{author}{\bibfnamefont{M.}~\bibnamefont{D'Elia}},
\bibinfo{author}{\bibfnamefont{F.}~\bibnamefont{Di}~\bibnamefont{Renzo}} 
\bibnamefont{and}
\bibinfo{author}{\bibfnamefont{M.}~\bibfnamefont{P.}~\bibnamefont{Lombardo}},  
\bibinfo{journal}{Phys. Rev.\ D} \textbf{\bibinfo{volume}{76}},
\bibinfo{pages}{114509} (\bibinfo{year}{2007}). 
%
\bibitem[{\citenamefont{Chen and Luo}(2005)}]{Chen}
\bibinfo{author}{\bibfnamefont{H.}~\bibfnamefont{S.}~\bibnamefont{Chen}} \bibnamefont{and}
\bibinfo{author}{\bibfnamefont{X.}~\bibfnamefont{Q.}~\bibnamefont{Luo}},  
\bibinfo{journal}{Phys.\ Rev.\  D} \textbf{\bibinfo{volume}{72}},
\bibinfo{pages}{034504} (\bibinfo{year}{2005}); 
\bibinfo{howpublished}{arXiv:hep-lat/0702025}; 
\bibinfo{author}{\bibfnamefont{L.}~\bibfnamefont{K.}~\bibnamefont{Wu}}, 
\bibinfo{author}{\bibfnamefont{X.}~\bibfnamefont{Q.}~\bibnamefont{Luo}},  
\bibnamefont{and}
\bibinfo{author}{\bibfnamefont{H.}~\bibfnamefont{S.}~\bibnamefont{Chen}}, 
\bibinfo{journal}{Phys.\ Rev.\  D} \textbf{\bibinfo{volume}{76}},
\bibinfo{pages}{034505} (\bibinfo{year}{2007}). 
%
\bibitem[{\citenamefont{Boyd et al.}(1996)}]{Boyd1}
\bibinfo{author}{\bibfnamefont{G.}~\bibnamefont{Boyd}},
\bibinfo{author}{\bibfnamefont{J.}~\bibnamefont{Engels}},
\bibinfo{author}{\bibfnamefont{F.}~\bibnamefont{Karsch}},
\bibinfo{author}{\bibfnamefont{E.}~\bibnamefont{Laermann}},
\bibinfo{author}{\bibfnamefont{C.}~\bibnamefont{Legeland}},
\bibinfo{author}{\bibfnamefont{M.}~\bibnamefont{L\"{u}tgemeier}},
\bibnamefont{and}
\bibinfo{author}{\bibfnamefont{B.}~\bibnamefont{Petersson}},
 \bibinfo{journal}{Nucl. Phys.} \textbf{\bibinfo{volume}{B469}},
\bibinfo{pages}{419} (\bibinfo{year}{1996}). 
%
\bibitem[{\citenamefont{Kaczmarek}(2002)}]{Kaczmarek1}
\bibinfo{author}{\bibfnamefont{O.}~\bibnamefont{Kaczmarek}},
\bibinfo{author}{\bibfnamefont{F.}~\bibnamefont{Karsch}},
\bibinfo{author}{\bibfnamefont{P.}~\bibnamefont{Petreczky}},
\bibnamefont{and}
\bibinfo{author}{\bibfnamefont{F.}~\bibnamefont{Zantow}},  
  \bibinfo{journal}{Phys. Lett.\ B} \textbf{\bibinfo{volume}{543}},
  \bibinfo{pages}{41} (\bibinfo{year}{2002}).
%
\bibitem[{\citenamefont{Karsch, Leermann and Peikert}(2001)}]{Karsch}
\bibinfo{author}{\bibfnamefont{F.}~\bibnamefont{Karsch}}, 
\bibinfo{author}{\bibfnamefont{E.}~\bibnamefont{Laermann}}, 
\bibnamefont{and}
\bibinfo{author}{\bibfnamefont{A.}~\bibnamefont{Peikert}},
\bibinfo{journal}{Nucl. Phys.} \textbf{\bibinfo{volume}{B605}},
\bibinfo{pages}{579} (\bibinfo{year}{2001}). 
%
\bibitem[{\citenamefont{Boyd et al.}(1995)}]{Boyd2}
\bibinfo{author}{\bibfnamefont{G.}~\bibnamefont{Boyd}}, 
\bibinfo{author}{\bibfnamefont{S.}~\bibnamefont{Gupta}}, 
\bibinfo{author}{\bibfnamefont{F.}~\bibnamefont{Karsch}}, 
\bibinfo{author}{\bibfnamefont{E.}~\bibnamefont{Laermann}}, 
\bibinfo{author}{\bibfnamefont{B.}~\bibnamefont{Petersson}}, 
\bibnamefont{and}
\bibinfo{author}{\bibfnamefont{K.}~\bibnamefont{Redlich}},
  \bibinfo{journal}{Phys. Lett.\ B} \textbf{\bibinfo{volume}{349}},
  \bibinfo{pages}{170} (\bibinfo{year}{1995}).
%
\bibitem[{\citenamefont{Kaczmarek et al.}(1995)}]{Kaczmarek2}
\bibinfo{author}{\bibfnamefont{O.}~\bibnamefont{Kaczmarek}},
\bibnamefont{and}
\bibinfo{author}{\bibfnamefont{F.}~\bibnamefont{Zantow}},  
  \bibinfo{journal}{Phys.\ Rev.\  D} \textbf{\bibinfo{volume}{71}},
  \bibinfo{pages}{114510} (\bibinfo{year}{2005}).
%
\bibitem[{\citenamefont{Aoki et al.}(2006)}]{Aoki}
\bibinfo{author}{\bibfnamefont{Y.}~\bibnamefont{Aoki}}, 
\bibinfo{author}{\bibfnamefont{Z.}~\bibnamefont{Fodor}}, 
\bibinfo{author}{\bibfnamefont{S.}~\bibnamefont{D.}~\bibnamefont{Katz}}, 
\bibnamefont{and}
\bibinfo{author}{\bibfnamefont{K.}~\bibnamefont{K.}~\bibnamefont{Szab\'{o}}}, 
  \bibinfo{journal}{Phys. Lett.\ B} \textbf{\bibinfo{volume}{643}},
  \bibinfo{pages}{46} (\bibinfo{year}{2006}).

\end{thebibliography}
\end{document}